\documentclass[]{emulateapj}

\shorttitle{Submillimeter Observations of Titan}
\shortauthors{Gurwell}

\begin{document}

\slugcomment{Accepted by ApJ Letters: July 7, 2004}

\def\httcn{H$^{13}$CN}
\def\hcftn{HC$^{15}$N}
\def\hctn{HC$_3$N}
\def\degree{$^\circ$}

\title{Submillimeter Observations of Titan:  Global Measures of
Stratospheric Temperature, CO, HCN, 
\hctn , and the
Isotopic Ratios $^{12}$C/$^{13}$C and $^{14}$N/$^{15}$N}

\altaffiltext{1}{Harvard-Smithsonian Center for Astrophysics, 
  60 Garden Street, Cambridge, MA 02138, USA}

\author{Mark A. Gurwell\altaffilmark{1}}
\email{mgurwell@cfa.harvard.edu}

\begin{abstract}

Interferometric observations of the atmosphere of Titan were performed
with the Submillimeter Array on two nights in February 2004 to
investigate the global average vertical distributions of several
molecular species above the tropopause.  Rotational transitions of CO,
isomers of HCN, and \hctn\ were simultaneously recorded.  The
abundance of CO is determined to be 51$\pm$4 ppm, constant with
altitude.  The vertical profile of HCN is dependent upon the assumed
temperature but generally increases from 30 ppb at the condensation
altitude ($\sim$83 km) to 5 ppm at $\sim$300 km.  Further, the central
core of the HCN emission is strong and can be reproduced only if the
upper stratospheric temperature increases with altitude.  The isotopic
ratios are determined to be $^{12}$C/$^{13}$C=132$\pm$25 and
$^{14}$N/$^{15}$N=94$\pm$13 assuming the Coustenis \& B\'ezard (1995)
temperature profile.  If the Lellouch (1990) temperature profile is
assumed the ratios decrease to $^{12}$C/$^{13}$C=108$\pm$20 and
$^{14}$N/$^{15}$N=72$\pm$9.  The vertical profile of \hctn~ is
consistent with that derived by Marten et al. (2002).

\end{abstract}


\section{Introduction}

The atmosphere of Titan is cold, dense and nitrogen-dominated.  Water,
CO, nitriles and hydrocarbons have been detected by Voyager spacecraft
and from Earth.  Understanding this unique atmosphere is
a major focus in planetology, particularly with the imminent arrival
of the Cassini spacecraft with the Huygens probe destined for Titan.
Studies of the photochemistry of Titan's atmosphere have been ongoing
since the Voyager era, and have become increasingly sophisticated
(e.g. Yung, Allen, \& Pinto 1984; Toublanc et al. 1995; Lara et
al. 1996; Lebonnois et al. 2001; Lebonnois et al. 2003).  These models
rely on accurate observational data as primary constraints. 

Trace molecules such as CO and nitriles contribute many emission
features that can be used to probe the vertical structure of its
atmosphere.  As access to the millimeter and submillimeter wavelength
bands has improved, there has been an series of ground-based
observational studies of Titan's atmosphere (e.g. Muhleman, Berge, \&
Clancy 1984; Tanguy et al 1990; Gurwell \& Muhleman 1995; Hidayat et
al. 1997 \& 1998; Gurwell \& Muhleman 2000; Marten et al 2002).  The
Submillimeter Array\footnote{ The Submillimeter
Array is a joint project between the Smithsonian Astrophysical
Observatory and the Academia Sinica Institute of Astronomy and
Astrophysics, and is funded by the Smithsonian Institution and the
Academia Sinica.}  (SMA) is well suited for observing Titan.

\section{Observations and Data Reduction}

We utilized the SMA to obtain high precision spectroscopy of the
atmosphere of Titan on 1 Feb 2004 and 21 Feb 2004. Parameters for
these observations are in Table 1; we briefly discuss here a few
important points.

The nearly 2 GHz total bandwidth/sideband allowed for several
molecular transitions to be observed.  The lower sideband (LSB)
includes the CO(3-2), \httcn (4-3), \hcftn (4-3), and \hctn (38-37)
rotational transitions.  The upper sideband (USB) covered both the
ground and vibrationally excited HCN(4-3) $\nu_2=0,1$ transitions, and
the \hctn (39-38) transition.

The large separation between Saturn and Titan relative to the primary
beam size severely downweighted the integrated flux of Saturn.  In
addition, Saturn is highly resolved on the baselines used, further
reducing any contribution from Saturn.  We consider any contamination
from Saturn is negligible.

Data from each night were independently calibrated to remove
instrumental gain drifts over time and to correct the complex bandpass
response.  Titan's signal strength was sufficient for application of
phase self-calibration.  The flux scale was initially determined using
relatively line free regions of the Titan spectrum.  In the absence of
molecular lines, the emission from Titan in the submillimeter band
comes from the region near the tropopause ($\sim$40 km, 71 K), and
this 'continuum' emission provided the primary flux calibration
reference.  Further adjustments to the flux scale were performed
during analysis.

Spatially unresolved (e.g. ``zero-spacing'') spectra for
each night were obtained by fitting the visibility data for
each channel with a model, correcting for the spatial sampling of the
interferometer. To minimize errors associated with this model fitting
only baselines less than 125 m (corresponding to angular scales
greater than 1.4$''$) were used.  The relative sideband scale in the
average spectra is better than 0.5\%.  The spectra from the two days
were scaled to reflect a common apparent size of 0.866$''$, then added
to increase the signal to noise ratio.  Figure 1 presents the final
calibrated spectra.

\begin{deluxetable}{lcc}
\tablecolumns{3}
\tablewidth{0pt}
\tablecaption{Titan Observational Parameters}
\tablehead{
\colhead{Parameter} & \multicolumn{2}{c}{Value} }
\startdata
LO Frequency             & \multicolumn{2}{c}{349.95 GHz}  \\
Total bandwidth/sideband & \multicolumn{2}{c}{1968 MHz}    \\
Channel spacing &  \multicolumn{2}{c}{0.8125 MHz} \\
Primary beam HPBW &  \multicolumn{2}{c}{$33''$}   \\
Synthesized resolution &  \multicolumn{2}{c}{$\sim 0.5''$} \\
Calibration sources & \multicolumn{2}{c}{Ceres, Callisto} \\
Observing dates & 01 Feb 2004 & 21 Feb 2004        \\
Number of antennas  &   5       &   6                  \\
Integration time   & 257 m  & 232 m            \\
Zenith atmospheric opacity\tablenotemark{a} & 0.18-0.28 & 0.15-0.20 \\
Titan diameter\tablenotemark{b} & 0.866$''$ & 0.841$''$ \\
Separation from Saturn  & $171''-182''$ & $113''-126''$ \\
Subearth Longitude & $233.4-240.5$ & $324.4-331.1$ \\
Subearth Latitude  &  -25.66 & -25.82          \\
\enddata
\tablenotetext{a}{at LO frequency}
\tablenotetext{b}{for surface; assumes radius of 2575 km}
\label{tab:obs}
\end{deluxetable}

\section{Analysis, Results, and Discussion}

The measured lineshape from a planetary atmosphere is a complex
function of the vertical profiles of temperature and absorber
abundances.  This information is ``encoded'' in the lineshape through
pressure broadening, and can be retrieved within certain bounds
through suitable numerical inversion or radiative transfer modeling.
We updated and expanded our iterative least squares inversion
algorithm, previously used to measure the atmospheric abundance of CO
on Titan (Gurwell \& Muhleman 1995; Gurwell \& Muhleman 2000), to
allow analysis of the rotational lines observed for this work.

The base thermal profile model was the equatorial profile A determined
by Coustenis and B\'ezard (1995) based upon Voyager 1/IRIS spectra of
methane and the Voyager radio occultation results; we refer to this as
profile A as well.\footnote{Also used in Hidayat et al.(1997, 1998),
Gurwell \& Muhleman (2000), and Marten et al (2002).  This model is
appropriate since the observations are of unresolved (whole-disk)
spectra, which are heavily weighted by emission from equatorial and
low southern latitudes.}  We investigated two ad hoc temperature
profiles related to profile A.  Profile B is isothermal at 180 K above
230 km, and profile C is isothermal at 180 K from 230 to 350 km, with
a positive temperature gradient above 350 km.  Profile D is an older
profile from a reanalysis of the Voyager radio occultation data
(Lellouch 1990; see Fig. 2a).  The base molecular abundance model has
CO uniformly mixed at 52 ppm (Gurwell \& Muhleman 2000), the HCN and
\hctn\ profiles following the results of Marten et al (2002), and
isotopic ratios of $^{12}$C/$^{13}$C=89 (Hidayat et al 1997) and
$^{14}$N/$^{15}$N=65 (Marten et al 2002).

\subsection{CO} 

CO has a long chemical lifetime in the atmosphere of Titan ($\sim10^9$
years; Yung, Allen \& Pinto 1984) and is expected to be uniformly
mixed from the surface to high altitude.  Observational evidence has
shown this to be the case, though there is disagreement regarding the
value (Gurwell \& Muhleman 1995; Hidayat et al 1998; Gurwell \&
Muhleman 2000).  Here we assumed that the CO profile is constant with
altitude, and iteratively fit the LSB spectrum for the stratospheric
CO abundance.

There is a modest error associated with the overall flux scale of the
spectrum, which was set using a basic model of the Titan spectrum in a
relatively line free region near 344.0 GHz.  This error was calibrated
by scaling the current model spectrum to best fit the observed
spectrum at each iterative step.  This essentially eliminates the
effect of an overall error in the flux density scale by forcing a best
fit to the normalized lineshape rather than to a fixed flux scale
spectrum.  This calibration is robust for interferometric
observations, which preserve the relative line to continuum of the
source emission.  The output of the iterative inversion is the
best-fit CO abundance, and a best-fit correction to the overall flux
scale.  This scale factor is applicable to both sidebands.

The best fit CO abundance is insensitive to the different temperature
models, with solutions of 50.6, 51.5, 51.8, and 52.0 ppm (all
$\pm$2.7) obtained with temperature models A, B, C, and D,
respectively.  Taking the uncertainty of the upper atmospheric
temperature, we consider all of these solutions as equally valid, and
adopt a mean stratospheric CO abundance of 51$\pm$4 ppm.  This agrees
well with the previous stratospheric CO measurements of Gurwell \&
Muhleman (2000; 52$\pm$6 ppm), Gurwell \& Muhleman (1995; 50$\pm$10
ppm), and Muhleman, Berge, \& Clancy (1984; 60$\pm$40 ppm), as well as
the initial measurement of CO on Titan from near IR observations
(Lutz, de Bergh, \& Owen, 1983; 48$^{+100}_{-32}$ in the troposphere).
However, these values are at odds with other recent measurments, such
as Hidayat et al. (1998; 27$\pm$5 ppm in the stratosphere), Noll et
al. (1996; 10$^{+10}_{-5}$ in the troposphere), and Lellouch et
al. (2003; 32$\pm$10 ppm in the troposphere).

\subsection{HCN: Abundance Profile}

HCN is expected to saturate in the cold lower stratosphere, with a
condensation altitude near 83 km (see Marten et al 2002).  Solutions
for the vertical profile of HCN determined assuming the four
temperature models are shown in Fig. 2b, along with the most recent
published HCN determination by Marten et al (2002).  Model
calculations are compared to the SMA data in Figs. 2c and 2d.

HCN profiles obtained assuming temperature models A, B, and C are
nearly identical, with an exponential increase from 30 ppb at the
condensation altitude to 5 ppm at $\sim$300 km.  The profile
corresponding to model D is similar but less abundant below 300 km.
All solutions are inconsistent with previous analyses showing a nearly
constant HCN abundance above 180-200 km (Hidayat et al. 1997, Marten
et al. 2002).  Radiative transfer modeling using the HCN profile of
Marten et al. (2002) results in a poor fit to the HCN line wings,
indicating that profile is overabundant in HCN below $\sim$180 km.
The nearly constant HCN profile above 180 km does not provide enough
contribution to the central $\pm$50 MHz of the HCN lineshape,
resulting in a poor fit to the line center (see Fig. 2c).  Our results
strongly favor a steadily increasing HCN mixing ratio from the
condensation altitude to at least 300 km.

\subsection{HCN: Upper Atmospheric Temperature} 

None of the calculations using temperature models A or D provide a
good fit to the HCN(4-3) line core, which exhibits strong emission in
the central 5 MHz.  We find no manipulation of the HCN mixing ratio
profile that can reproduce this line core assuming these standard
temperature profiles.

Yelle and Griffith (2003) present a reanalysis of 3$\mu$m HCN
fluorescence observations obtained by Geballe et al (2003).  They find
the observations can be fit assuming the HCN profile determined by
Marten et al (2002) up to 450 km and a temperature profile similar to
model A.  Their temperature model (from Yelle et al. 1997) was
constructed to satisfy a variety of observational and theoretical
constraints.  In relation to this work, a key component of this
temperature profile is a smooth decrease from about 180 K near 250 km
(250 $\mu$bar) to a mesopause temperature of $\sim$135 K near 600 km
(0.2 $\mu$bar), a trait shared with model A below 450 km.

The HCN spectrum presented here is clearly incompatible with the Yelle
et al (1997) and model A temperature profiles.  A negative temperature
gradient above 250 km results in a line core in absorption (see
Fig. 2d).  An isothermal upper atmosphere (model B) can account for
some of the strong core emission, but not all.  The progression of
model fits shown in Fig. 2d suggests that the only way to fit the
intense line core is to have an upper stratosphere that is {\it
warmer} than the stratosphere near 250 km.  Model C has such a profile
and provides the closest fit to the data.

There is little observational evidence for a temperature maximum
occuring only near 250 km (250 $\mu$bar).  The thermal profile below
180 km (1 mbar) is measured from IR and radio observations
(e.g. Coustenis et al. 2003), and the thermospheric temperatures are
constrained by UV occultation measurements (Smith et al. 1983).
Several lines of evidence point to a mesopause around 600 km, above 1
$\mu$bar.  Thermal profiles determined from observations of the
occultation of star 28 Sgr by Titan do sense the 300-500 km altitude
range (Hubbard et al 1993), and especially for the southern hemisphere
are relatively consistent with an isothermal atmosphere below
$\sim$350 km, though the large scatter in the inversion results for
different data sets makes a definitive statement difficult, and a
positive temperature gradient from 300 to 350 km is possible.

We propose here that the strong HCN emission core is solid evidence
for a significantly warmer atmosphere above 250 km compared to the
profiles from Yelle et al (1997) and Coustenis et al (2003).  The
possibility of a mesopause above 1 $\mu$bar is not precluded, but we
find that the profile likely cannot have a negative temperature
gradient below roughly 450 km.

\subsection{Isotopic Ratios} 

Iterative solutions for the ratios of HCN/\httcn~ and HCN/\hcftn~ were
obtained assuming the less abundant isomers had the same vertical
profile shape as HCN, and that the isomer ratios reflect the overall
ratios of $^{12}$C/$^{13}$C and $^{14}$N/$^{15}$N in the atmosphere.
The solutions we obtain show that retrievals of \httcn~ and \hcftn~
are less sensitive than that of HCN to the assumed temperature profile
below 300 km.  Therefore, there is a dependence of the isotopic ratios
on the assumed temperature profile.  For models A, B, and C (which
are identical below 230 km) we find $^{12}$C/$^{13}$C=132$\pm$25 and
$^{14}$N/$^{15}$N=94$\pm$13.  Both ratios decrease assuming model D to
$^{12}$C/$^{13}$C=108$\pm$20 and $^{14}$N/$^{15}$N=72$\pm$9.

Terrestrial values for the carbon and nitrogen isotopic ratios are 89
and 272, respectively (Fegley 1995).  Our observations are consistent
with nitrogen being enriched in the heavier isotope by a factor of 2.9
and carbon being depleted in the heavier isotope by a factor of 0.5,
relative to terrestrial values.  In contrast, using model A, Hidayat
et al (1997) derived $^{12}$C/$^{13}$C=95$\pm$25 and Marten et al
(2002) derived $^{14}$N/$^{15}$N=65$\pm$5.  These previously measured
isotopic ratios are consistent with results we obtain assuming
temperature profile D, but are not consistent with results for
profiles A, B, and C. 

\subsection{HC$_3$N} 

Previous observations of \hctn\ have shown it is most abundant above
300 km (Marten et al 2002).  The observations presented here are
unresolved at a resolution of 0.8125 MHz, which is consistent with an
abundance limited to altitudes above 300 km.  Since we did not resolve
the lineshapes retrieval of a profile was not practical.  We instead
modeled the overall column abundance assuming that the shape of the
profile determined by Marten et al (2002) was correct.  We find that
the \hctn\ transitions we observed are consistent with the Marten et
al. (2002) profile for all the model temperature profiles.

\acknowledgements

We thank all members of the SMA Team that made these observations
possible. We thank David Wilner for comments on several drafts of the
paper, and to the anonymous reviewer who pointed out several relevant
papers from the past year.  We extend special thanks to those of
Hawaiian ancestry on whose sacred mountain we are privileged to be
guests.

\clearpage

\begin{figure}
\epsscale{1.15}
\rotatebox{0}{
\plotone{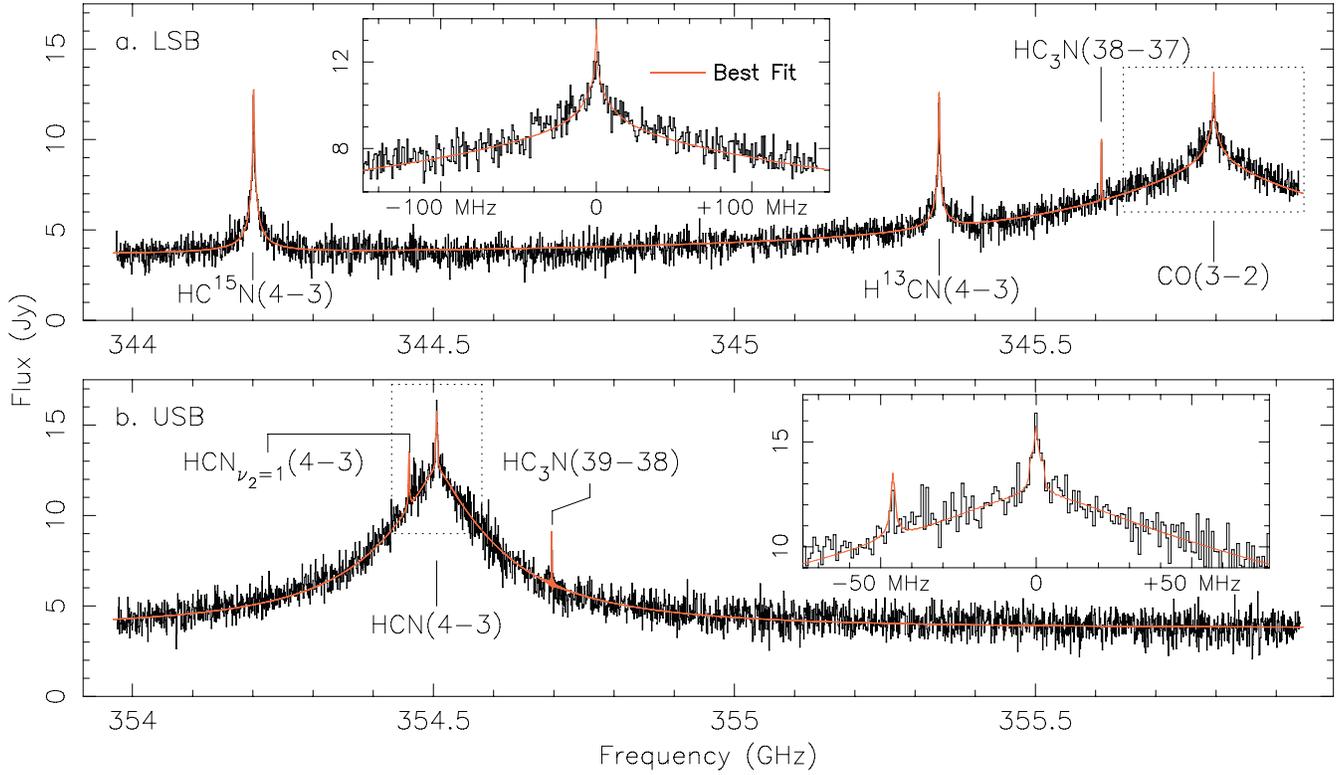}}
\caption{Submillimeter whole-disk spectrum of Titan obtained with the
SMA, with an overlayed best fit model spectrum. (a) Lower sideband
containing CO(3-2), \httcn (4-3), \hcftn (4-3), and \hctn (38-37)
rotational transitions.  (b) Upper sideband with both the ground and
vibrationally excited HCN(4-3) $\nu_2=0,1$ transitions, and the \hctn
(39-38) transition.  }

\end{figure}

\vskip 0.75 in

\begin{figure}
\epsscale{1.15}
\rotatebox{0}{
\plotone{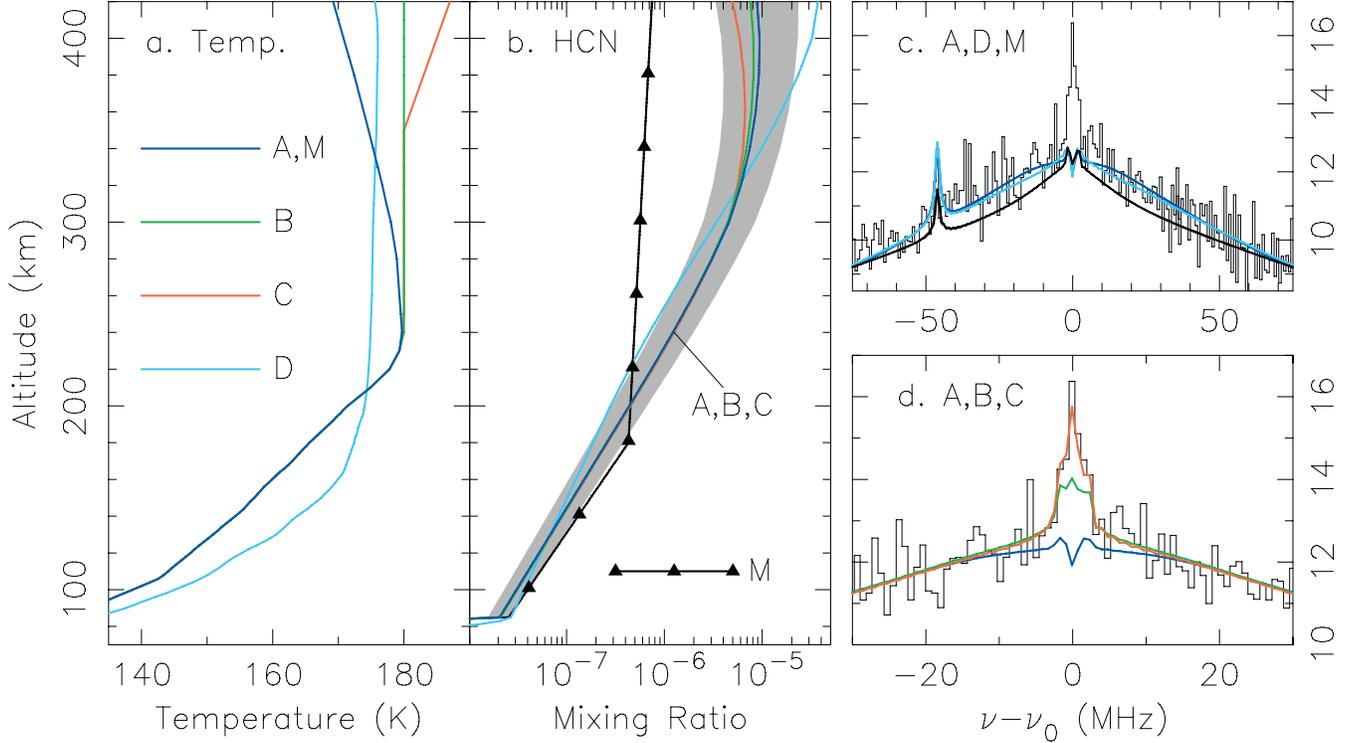}}
\caption{(a) Model temperature profiles used in inversion analysis.
(b) HCN best fit profile solutions assuming temperature models A, B,
C, and D, and the HCN profile (M) determined by Marten et
al. (2002). Grey scale represents the error envelope for solutions A,
B, and C.  (c) Observed and best fit HCN line center spectra for HCN
profiles A, D, and M.  (d) Observed and best fit HCN line core spectra
for HCN profiles A, B, C.  }
\end{figure}

\end{document}